\documentclass[conference]{IEEEtran}
\IEEEoverridecommandlockouts
\usepackage{amsmath,amssymb,amsfonts}
\usepackage{algorithmic}
\usepackage{graphicx}
\usepackage{textcomp}
\usepackage{xcolor}

\usepackage{calc}
\usepackage{enumitem}
\usepackage{xspace}

\usepackage{url}

\usepackage[numbers,square,sort&compress]{natbib}
\usepackage{balance}

\usepackage{booktabs}
\usepackage{tabularx}
\usepackage{multirow}

\usepackage{microtype}

\renewcommand{\paragraph}[1]{{\vskip 8pt \noindent\bf #1 }}

\newcommand{\srcimg}{\ensuremath{S}\xspace}
\newcommand{\tarimg}{\ensuremath{T}\xspace}
\newcommand{\attimg}{\ensuremath{A}\xspace}
\newcommand{\outimg}{\ensuremath{D}\xspace}
\newcommand{\deltaS}{\ensuremath{\Delta}\xspace}
\newcommand{\scalefunc}{\ensuremath{\mathrm{scale}}}

\newcommand{\ti}{\ensuremath{Z}\xspace}

\newcommand{\goalA}{O1\xspace} %
\newcommand{\goalB}{O2\xspace} %

\definecolor{goeblue}{RGB}{0,51,102}
\definecolor{denceblue}{RGB}{6,107,176}
\definecolor{sunygreen}{RGB}{0,166,77}
\definecolor{darkgreentone}{RGB}{60,179,113}
\definecolor{tubsredSec}{cmyk}{0.0,1.00,0.6,0.6}
\definecolor{tubsredPrim}{cmyk}{0.1,1.0,0.8,0.0}

\def\BibTeX{{\rm B\kern-.05em{\sc i\kern-.025em b}\kern-.08em
    T\kern-.1667em\lower.7ex\hbox{E}\kern-.125emX}}
\begin{document}

\title{Backdooring and Poisoning Neural Networks\\with 
	Image-Scaling Attacks\thanks{Published at {IEEE} Deep Learning and
		Security Workshop (DLS) 2020, co-located with the 41st {IEEE} 
		Symposium on Security and Privacy (S\&P)}}

\author{
	{\rm Erwin Quiring and 
		Konrad Rieck}\\[1mm]
	\begin{minipage}{8cm} 
		\centering \it
		Technische Universit\"at Braunschweig, Germany
	\end{minipage} 	  %
}

\maketitle

\begin{abstract}
Backdoors and poisoning attacks are a major threat to the security of 
machine-learning and vision systems. Often, however, these attacks 
leave visible artifacts in the images that can be visually detected and 
weaken the efficacy of the attacks.  In this paper, we propose a novel 
strategy for hiding backdoor and poisoning attacks. Our approach builds 
on a recent class of attacks against image scaling. These attacks 
enable manipulating images such that they change their content when 
scaled to a specific resolution. By combining poisoning and 
image-scaling attacks, we can conceal the trigger of backdoors as well 
as hide the overlays of clean-label poisoning. Furthermore, we consider 
the detection of image-scaling attacks and derive an adaptive attack. 
In an empirical evaluation, we demonstrate the effectiveness of our 
strategy.  First, we show that backdoors and poisoning work equally 
well when combined with image-scaling attacks. Second, we demonstrate 
that current detection defenses against image-scaling attacks are 
insufficient to uncover our manipulations. Overall, our work provides a 
novel means for hiding traces of manipulations, being applicable to 
different poisoning approaches.
\end{abstract}

\vspace{0.2cm}

\section{Introduction}
Machine Learning is nowadays used in various security-critical 
applications that range from intrusion detection and medical 
systems to autonomous cars. Despite remarkable results, 
research on the security of machine learning has revealed various 
possible attacks.
A considerable threat are poisoning attacks during the training 
process~\citep[e.g.][]{BigNelLas11, GuDolGar17, LiuMaAaf+18}. 
Deep learning applications usually require a large number of 
training instances, so that there is a risk of an insider carefully 
manipulating a portion of the training data. Moreover, the training 
process can be outsourced either due to a lack of expertise in deep 
learning or due to missing computational resources to train large 
networks---again giving the chance to manipulate training data and the 
model.

In the context of deep learning, recent research has demonstrated that 
neural networks can be modified to return targeted responses without an 
impact on their behavior for benign inputs. An adversary, for instance, 
can insert a pattern in some training images of a particular 
target class, so that the network learns to associate the pattern with 
this class. If the pattern is added to arbitrary images, the network 
returns the target class.
However, a major drawback of most attacks is the visibility of data 
manipulations either at training or test time~\citep{GuDolGar17, 
LiuMaAaf+18}. The attack is thus revealed if human beings 
audit the respective images. 

\citet{XiaCheShe+19} have recently presented a novel attack 
vulnerability in the data preprocessing of typical machine learning 
pipelines. An adversary can slightly manipulate an image, such that an
\emph{image scaling} algorithm produces a novel and unrelated image in 
the network's input dimensions. The attack exploits that images are 
typically larger than the input dimensions and thus need to be scaled.

\begin{figure}[t]
	\centering
	\includegraphics{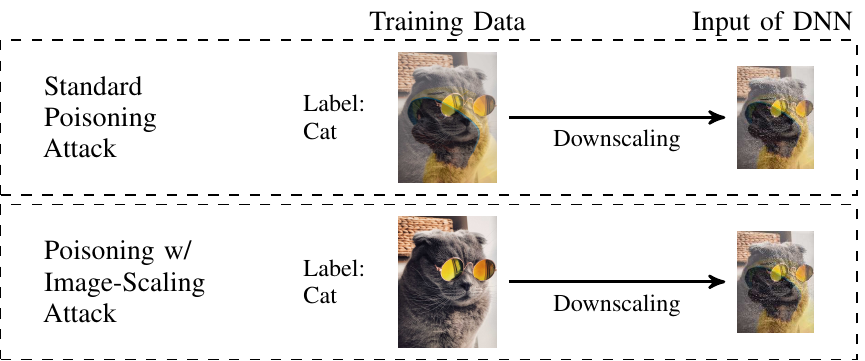}
	\vspace{-0.59cm}
	\caption{Example of a clean-label poisoning 
	attack~\citep{ShaHuaNaj+18}: a neural network 
	learns to classify a dog as cat by blending the dog with multiple 
	cat images. %
	Image-scaling attacks allow more insidious poisoning attacks.
	The dog as manipulation is not visible in the training data and 
	appears only after downscaling.
	}
	\label{fig:intro_example_poisoning}
\end{figure}

This novel attack directly addresses the shortcomings of 
most poisoning attacks by allowing an adversary to conceal data 
manipulations. As an example, Figure~\ref{fig:intro_example_poisoning} 
shows a clean-label poisoning attack~\citep{ShaHuaNaj+18} on the 
popular TensorFlow library. The network will learn to classify 
the dog as cat if this dog is repeatedly inserted into 
varying images showing cats during training.
In the attack's standard version, the slight manipulation of the 
training image is still noticeable. Yet, image-scaling attacks conceal 
the manipulation of the training data effectively. The dog appears only 
in the downscaled image which is finally used by the neural network.

This paper provides the first analysis on the combination of data 
poisoning and image-scaling attacks. Our findings show that an 
adversary can significantly conceal image manipulations of current 
backdoor attacks~\citep{GuDolGar17} and clean-label 
attacks~\citep{ShaHuaNaj+18} without an impact on their overall attack 
success rate.
Moreover, we demonstrate that defenses---designed to detect 
image-scaling attacks---fail in the poisoning scenario.
We examine the histogram- and color-scattering-based detection as 
proposed by~\citet{XiaCheShe+19}. In an empirical evaluation, we show 
that both defenses cannot detect backdoor attacks due to bounded, 
local changes. We further derive a novel adaptive attack that 
significantly reduces the performance of both defenses in the 
clean-label setting.
All in all, our findings indicate a need for novel, robust detection 
defenses against image-scaling~attacks.
\paragraph{Contributions.}  In summary, we make the following
contributions in this paper:
\begin{itemize}  \setlength{\itemsep}{3pt}
	
	\item \emph{Combination of data poisoning and image-scaling 
	attacks.}  
	We provide the first analysis on poisoning attacks that are 
	combined with image-scaling attacks. We discuss two realistic 
	threat models and consider backdoor attacks as well as clean-label 
	poisoning attacks.
		
	\item \emph{Evaluation of defenses} 
	We evaluate current detection methods against image-scaling 
	attacks and show that backdoor attacks cannot be detected.
	
	\item \emph{Adaptive Attack.} We derive a novel variant of
	image-scaling attack that reduces the detection rate 
	of current scaling defenses. Our evaluation 
	shows that clean-label attacks cannot be reliably detected anymore.
\end{itemize}

The remainder of this paper is organized as follows: 
Section~\ref{sec:background} reviews the background of data poisoning 
and image-scaling attacks. Section~\ref{sec:poisoning-scaling} examines 
their combination with the respective threat scenarios and our adaptive 
attack.
Section~\ref{sec:eval} provides an empirical evaluation of attacks and 
defenses.
Section~\ref{sec:limitations} and~\ref{sec:relatedwork} present 
limitations and related work, respectively. 
Section~\ref{sec:conclusion} concludes the paper.

\section{Background}\label{sec:background}
Let us start by briefly examining poisoning and image-scaling 
attacks on machine learning. Both attacks operate at different stages 
in a typical machine learning pipeline and allow more powerful attacks 
when combined, as we will show in the remainder of this work.

\subsection{Poisoning Attacks in Machine Learning}
In machine learning, the training process is one of the most 
critical steps due to the impact on all subsequent applications. At 
this stage, poisoning attacks allow an adversary to change the overall 
model behavior~\citep[e.g.][]{KloLas10a,BigNelLas11}
or to obtain targeted responses for specific 
inputs~\citep[e.g.][]{GuDolGar17,ShaHuaNaj+18,LiuMaAaf+18} by 
manipulating the training data or learning model. Such attacks need to 
be considered whenever the training process is outsourced or an 
adversary has direct access to the data or model as 
insider~\citep{Sto15}. Moreover, a possible manipulation needs to be 
considered if a learning model is continuously updated with external 
data.

In this work, we focus on poisoning attacks against deep neural 
networks where the adversary manipulates the training data to obtain 
targeted predictions at test time. While particularly effective with a 
few changed training instances, most methods have the major shortcoming 
that the manipulation is visible~\citep[e.g.][]{GuDolGar17, 
	LiuMaAaf+18}. As a result, the attack can be easily uncovered if 
	the dataset is, for instance, audited by human beings.
We present two representative 
poisoning attacks in Section~\ref{sec:poisoning-scaling} and show 
that they can be easily combined with image-scaling attacks to
conceal manipulations significantly.

\begin{figure}[t]
	\centering
	\includegraphics{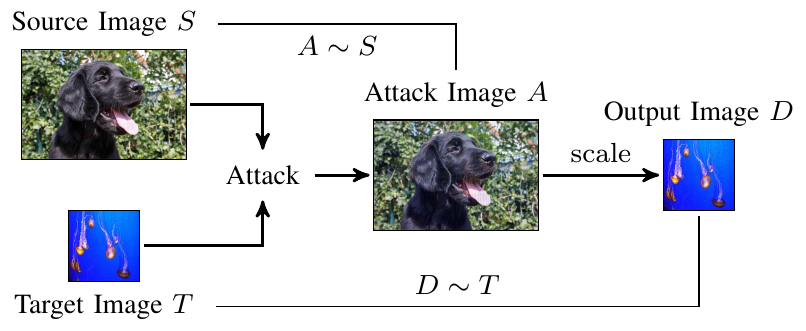}
	\vspace{-0.5em}
	\caption{Principle of image-scaling attacks: An
		adversary computes \attimg such that it looks like \srcimg but
		downscales to \tarimg.}
	\label{fig:attack_example}
\end{figure}

\subsection{Image-Scaling Attacks}
\label{subsec:scalingattacksbackground}
The preprocessing stage in a typical machine learning pipeline
is another critical point, surprisingly overlooked by previous 
work so far. \citet{XiaCheShe+19} have recently identified an attack 
possibility in the scaling routine of common machine learning 
frameworks. The attack exploits that most learning-based models expect 
a fixed-size input, such as $224 \times 224$ pixels for 
VGG19 and $299 \times 299$ pixels for 
InceptionV3~\citep{SimZis14,SzeVanIof+15}.
As images are usually larger than the input dimensions of learning 
models, a downscaling operation as preprocessing stage is mandatory. 
In this case, an adversary can slightly modify an image such that it 
changes its content after downscaling. She can thus create targeted 
inputs for a neural network being invisible in the original 
resolution before,~as~exemplified~by~Figure~\ref{fig:attack_example}.

\paragraph{Attack.}
In particular, the adversary slightly modifies a source image~\srcimg 
such that the resulting attack image~\mbox{\attimg = \srcimg + \deltaS} 
matches a target image~\tarimg after scaling. The attack can be modeled 
as the following quadratic optimization problem:
\begin{align}
&\min ( \Vert \deltaS \Vert_2^2 ) \; \;
\mathrm{s.t.} \; \; \Vert \scalefunc(\srcimg + \deltaS) - \tarimg
\Vert_{\infty} \leqslant \epsilon  \; .
\label{eq:opti_problem_basic}
\end{align}
Moreover, each pixel of \attimg needs to stay in the range of 
$[0,255]$ for 8-bit images. 
Note that an image-scaling attack is successful only if the 
following two goals are fulfilled:
\begin{description}[leftmargin=!,labelwidth=\widthof{\textit{(O2)}}]
	\item[\textit{(\goalA)}] The downscaled output~\outimg of the 
	attack image~\attimg is close to 
	the target image:  $\outimg \sim \tarimg $.
	\item[\textit{(\goalB)}]
	The attack image \attimg needs to be indistinguishable from the 
	source image:~\mbox{$\attimg \sim \srcimg$}.
\end{description}
For a detailed root-cause analysis of image-scaling attacks, we refer 
the reader to \citet{QuiKleArp20}.

\begin{figure}[h]
	\centering
	\includegraphics{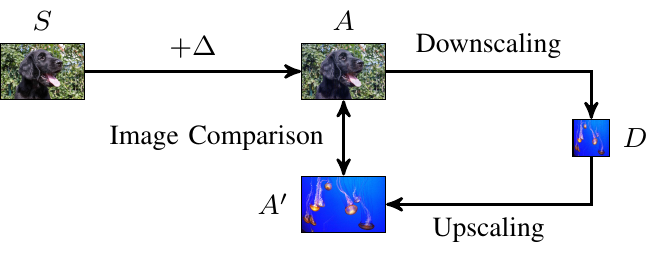}
	\vspace{-0.4em}
	\caption{Defense based on down- and upscaling with image comparison.
		The downscaled version~\outimg of \attimg is upscaled again and
		compared with \attimg.}
	\label{fig:defenses_detection_example}
\end{figure}

\paragraph{Detection.}
Two methods have been proposed to \emph{detect} image-scaling 
attacks~\citep{XiaCheShe+19}, that is, decide if an image was 
manipulated to cause another result after downscaling.
Both rest on the following idea, exemplified by 
Figure~\ref{fig:defenses_detection_example}: The downscaling of \attimg 
creates a novel image~\outimg which is unrelated to the original 
content from \attimg. If we upscale~\outimg back to its original 
resolution, we can compare \attimg and the upscaled 
version~$\attimg'$. In the case of an attack, both images will be 
different to each other.

The first method uses an \emph{intensity histogram} that counts 
the number of pixels for each value in the dynamic range of an image. 
To this end, a color image is converted into a grayscale
image before computing its histogram. The result is a 256~dimensional 
vector $v^h$ for 8-bit images.
The attack detection is now based on the cosine similarity 
between $\attimg$ and $\attimg'$: $s^h = \cos (v^h_1, v^h_{2} )$. A 
low score indicates an attack, as the distribution of both inputs do 
not match to each other.

The second method based on \emph{color scattering} considers spatial 
relations in an image. The color image is again converted to 
grayscale, and the average distance to the image center over 
all pixels with the same value is calculated, respectively. This forms 
a 256~dimensional vector~$v^s$. The respective vectors from $\attimg$ 
and $\attimg'$ are also compared by using the cosine similarity.

We finally note that defenses exist to \emph{prevent} 
image-scaling attacks~\citep[see][]{QuiKleArp20}. 
In contrast to detection, prevention blocks the attack from 
the beginning, but would not uncover that the dataset was 
manipulated, which is the focus in this~work.

\section{Data Poisoning Using Image-Scaling}
\label{sec:poisoning-scaling}
Equipped with an understanding of data poisoning and image-scaling 
attacks, we are ready to examine novel attacks.
The adversary's goal is that the model returns targeted predictions for 
specially crafted inputs while behaving normally for 
benign inputs. As image-scaling attacks provide a new means for 
creating novel content after downscaling, they are a perfect 
candidate to create less visible poisoning attacks.
We start by describing two plausible threat models and continue with a 
description of two attack variants.

\subsection{Two Realistic Threat Models}

\paragraph{Stealthiness during test time.}
It is common practice to outsource the training of large deep 
neural networks, either due to the lack of computational resources or 
due to missing expertise. 
In this scenario, an adversary can arbitrarily change the training data 
(or the model), as long as the returned model has the expected accuracy 
and architecture. The application of image-scaling attacks to hide 
changes is here not necessary. However, common backdoor attacks add 
visible triggers in test time instances to obtain a targeted 
prediction~\citep[e.g.][]{GuDolGar17,LiuMaAaf+18}. If such instances 
are examined, a visible backdoor, for instance, would directly reveal 
that a model has been tampered. These attacks can thus benefit from 
image-scaling attacks at test time.

\paragraph{Stealthiness during training time.} 
In the second scenario, the adversary has only access to the training 
data, but the training process or model cannot be manipulated. This 
scenario is particularly relevant with insiders who already have 
privileged access within a company~\citep{Sto15}.
Most poisoning attacks leave visible traces in the training data, so 
that the attack is detectable if audited by human beings. 
Consequently, image-scaling attacks are also useful for these scenarios.

\vspace{0.5em}
Finally, the application of image-scaling attacks
requires (a) knowledge of the used scaling algorithm and (b) the input 
size of the neural network. Their knowledge is plausible to assume if 
the attacker is an insider or trains the model herself.

\subsection{Enhanced Poisoning Attacks}
\label{subsec:poisoningwithimagescaling}
We study two representative poisoning attacks against deep neural 
networks: \emph{backdoor} and \emph{clean-label} attacks.
Both enable us to examine different approaches to manipulate images and 
their impact on image-scaling attacks and defenses.

\paragraph{Backdoor attack.}
As first attack, we use the \emph{BadNets} backdoor method 
from~\citet{GuDolGar17}. The adversary chooses a target label and a 
small, bounded backdoor pattern. This pattern is added to a limited 
number of training images and the respective label is changed to the 
target label. In this way, the classifier starts to associate this 
pattern with the target class.

We consider both threat models for the attack. As first variant, the 
adversary hides the poisoning on test time instances only. Thus, 
we use the BadNets method in its classic variant during the training 
process.
At test time, the adversary applies an image-scaling attack. The 
original image without backdoor represents the source image~\srcimg, 
its version with the backdoor in the network's input dimensions is the 
target image~\tarimg. By solving Eq.~\eqref{eq:opti_problem_basic}, the 
adversary obtains the attack image~\attimg that is passed to the 
learning system.
The pattern is only present after downscaling, so that an adversary can 
effectively disguise the neural network's backdoor.

In addition, we study the threat scenario where the adversary hides the 
modifications at training time. We use the same 
attack principle as before, but apply the image-scaling attack for the 
backdoored training images as well. 
This scenario is especially relevant if the backdoor is implemented in 
the physical world, e.g.\ on road signs. The trigger can be disguised 
in the training data by using image-scaling attacks, and easily 
activated in the physical world at test time (without a scaling attack).

\paragraph{Clean-label poisoning attack.}
As second attack, we consider the poisoning attack at training time as 
proposed by~\citet{ShaHuaNaj+18}. The attack does not change the label 
of the modified training instances. As a result, this poisoning 
strategy becomes more powerful in combination with image-scaling 
attacks: The manipulated images keep their correct class label and 
show no obvious traces of manipulation.

In particular, the adversary's objective is that the model 
classifies a specific and unmodified test set instance $\ti$ as a 
chosen target class $c_t$. To this end, the adversary chooses a set of 
images $X_i$ from $c_t$. Similar to watermarking, she embeds a 
low-opacity version of \ti into each image:
\begin{align}
X_i' = \alpha \cdot \ti + (1-\alpha) \cdot X_i .
\end{align}
If the parameter $\alpha$, for instance, is set to 0.3, features of 
$\ti$ are blended into $X_i$ while the manipulation is less 
visible. 
For an image-scaling attack, the adversary chooses $X_i$ as 
source image~\srcimg, and creates $X_i'$ as respective target 
image~\tarimg in the network's input dimensions. The computed attack 
image~\attimg serves as training image then.
The changed images are finally added to the training set together with 
their correct label $c_t$. As a result, the classifier learns to 
associate $\ti$ with $c_t$. At test time, $\ti$ can be passed to the 
learning system without any changes and is classified as~$c_t$. 
This attack enables us to study the detection of image-scaling attacks, 
if the entire image is slightly changed instead of adding a small and 
bounded trigger.

\subsection{Adaptive Image-Scaling Attack}
To hide poisoned images from detection, we additionally introduce a new 
variant of image-scaling attack. In particular, it targets the 
histogram-based defense, but is also effective against the 
color-scattering-based approach.

The difficult part is to create an attack image~\attimg that changes 
its appearance to~\tarimg after downscaling, but has a similar 
histogram if upscaled again, denoted as $\attimg'$.
To this end, we use the following strategy: we upscale the 
target image~\tarimg and perform a histogram matching to the source 
image~\srcimg. After slightly denoising the result to make the adjusted 
histogram smoother, we downscale the adapted image which gives us 
$\tarimg'$. We finally mount the image-scaling attack with \srcimg 
as source image and $\tarimg'$ as target. Although the content 
changes after down- and upscaling, the histogram remains similar.

\begin{figure}
	\centering
	\includegraphics{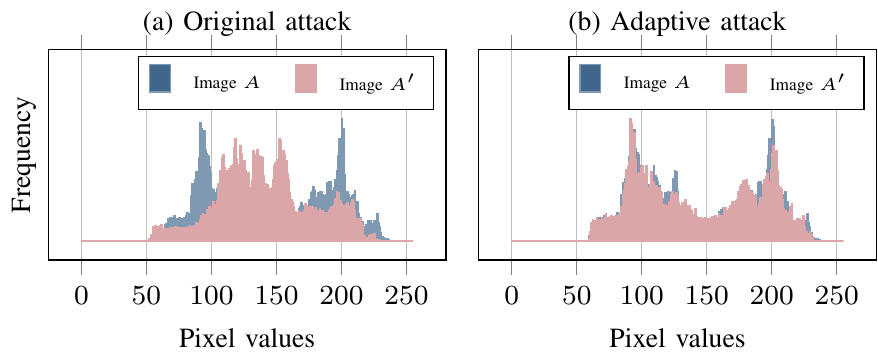}
	\includegraphics{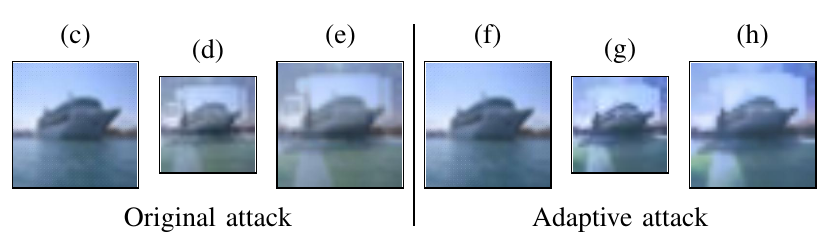}
	\vspace{-0.30cm} %
	\caption{Example of our adaptive image-scaling attack. 
		Plot (a) and (b) show the compared histograms for the original 
		and 
		our adaptive attack.
		Plot (c) and (f) show \attimg by using the 
		original attack and our adaptive version, respectively. Plot 
		(d) 
		and (g) show their respective downscaled version as input for 
		the 
		neural network, (e) and (h) the respective upscaled version 
		$\attimg'$.}
	\label{fig:adaptive_attack_example}
\end{figure}

Figures~\ref{fig:adaptive_attack_example}(a) and 
\ref{fig:adaptive_attack_example}(b) show an example with the 
histograms of $\attimg$ and $\attimg'$ for the original 
attack and our adapted attack, respectively. Our adaptive attack 
enables aligning the histograms of $\attimg$ and $\attimg'$ although 
both are visually different to each other, as depicted by
Figures~\ref{fig:adaptive_attack_example}(f) and (h). Moreover, the 
visual differences to the original attack are marginal.

However, the previous example also underlines that we do not obtain an 
exact histogram matching. The attack chances increase if source- and 
target image already have a similar color tone. Therefore, we let the 
adversary select the most suitable images for the attack. She mounts 
the attack on a larger number of adapted images and select those with 
the highest score.

\section{Evaluation}\label{sec:eval}
We continue with an empirical evaluation and perform the 
following experiments:
\begin{enumerate} \setlength{\itemsep}{2pt}
	\item \emph{Poisoning \& image-scaling attacks}. We first 
	demonstrate that poisoning attacks benefit from image-scaling 
	attacks. The attack performance remains constant while the data 
	manipulation is hard to notice.
	\item \emph{Detection defenses.} We demonstrate that currently 
	proposed defenses against image-scaling attacks cannot detect
	backdoor attacks.
	\item \emph{Adaptive attack.} We also show that clean-label attacks 
	cannot be detected if our new adaptive attack is applied to the 
	manipulated images.
	
\end{enumerate}

\subsection{Dataset \& Setup}
For our evaluation, we use the CIFAR-10 dataset~\citep{KriHin09}. Its 
respective default training set is further separated into a training 
(40,000 images) and validation set (10,000 images) that are used for 
model training. We choose the model architecture from \citet{CarWag17} 
which is commonly used in the adversarial learning literature. The 
model expects input images of size $32 \times 32 \times 3$. This 
simple configuration of dataset and model allows us to train a neural 
network for a variety of different attack configurations in feasible 
time. 

We implement the image-scaling attack in the strong variant as proposed 
by~\citet{XiaCheShe+19}, and set $\epsilon=1.0$ in 
Eq.~\eqref{eq:opti_problem_basic}. 
We use TensorFlow~(version~1.13) and report results for bilinear 
scaling, which is the default algorithm in TensorFlow. Due to our 
controlled scaling ratio, other scaling algorithms work identically and 
thus are omitted~\citep[see][]{QuiKleArp20}.

To evaluate image-scaling attacks realistically, we need source images 
in a higher resolution. To this end, we consider common scaling ratios 
from the ImageNet dataset~\citep{RusDenSu+15}. Its images are 
considerably larger than the input sizes of popular models for this 
dataset. VGG19, for instance, expects images with size $224 \times 224 
\times 3$~\citep{SimZis14}. Based on these results, 
we upscale the CIFAR-10 images to a size of $256 \times 256 \times 
3$ by using OpenCV's Lanczos algorithm. This avoids side effects if 
the same algorithm is used for upscaling and downscaling during an 
image-scaling attack and model training\footnote{If we use the same 
algorithm, we obtained even better results against image-scaling 
defenses, which might not be realistic with real-world images.}.

\subsection{Backdoor Attacks}
Our first experiment tests whether image-scaling attacks can 
effectively conceal backdoors. For a given target class, we
embed a filled black square in the lower-left corner as backdoor into 
training images. We perform the experiments for each class, 
respectively. To assess the impact of backdooring on benign inputs, we 
evaluate the accuracy on the unmodified CIFAR-10 test set. When 
evaluating backdoors on the test set, we exclude images from the target 
class. We report averaged results over the ten target 
classes. For each experiment, a baseline is added if the backdoor 
attack is applied without using an image-scaling attack.

\begin{figure}
	\centering
	\includegraphics{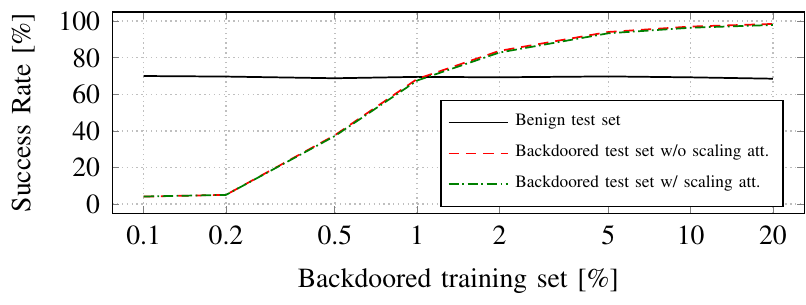}
	\vspace{-0.30cm}
	\caption{Backdoor attacks: 
		Percentage of obtained target classes on the 
		backdoored test 
		set, with and without image-scaling attacks for hiding the 
		backdoor. Scaling attacks have no negative impact on the 
		attack's success rate.
}
	\label{fig:eval_backdoor_plain_testtime}
\end{figure}
\begin{figure}
	\centering
	\includegraphics{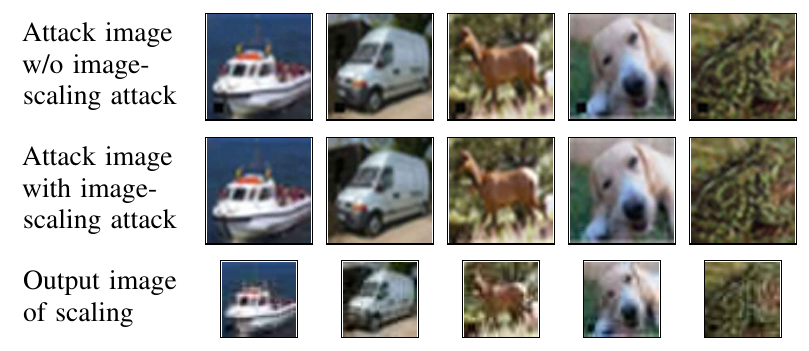}
	\vspace{-0.15cm}
	\caption{Backdoor attack examples.
The first and second row result in the third row after downscaling.
However, the second row relies on image-scaling
attacks and better hides the backdoor trigger.
}
	\label{fig:eval_backdoor_plain_testtime_examples}
\end{figure}

\paragraph{Attack performance.}
Figure~\ref{fig:eval_backdoor_plain_testtime} 
presents the success rate of the original attack for a varying number 
of backdoored training images. The adversary can successfully control 
the prediction by embedding a backdoor on test time instances. If 
5\% of the training data are changed, she obtains an almost perfect 
attack result. The test set accuracy with unmodified images does not 
change considerably.

The application of image-scaling attacks on the backdoored test time 
instances has no negative impact on the success rate.
At the same, the adversary can considerably hide the backdoor in 
contrast to the original attack, as 
Figure~\ref{fig:eval_backdoor_plain_testtime_examples} shows.
Although the backdoor's high contrast with neighboring pixels and its 
locality creates rather unusual noise in the backdoor area, the 
detection is hard if only quickly audited.

In addition, we evaluate the variant where the image-scaling attack is 
also applied on the \emph{training} data to hide the backdoor pattern. 
We obtain identical results to the previous scenario regarding the 
success rate and visibility of the backdoor pattern. 
In summary, image-scaling attacks considerably raise the bar 
to detect backdoors.

\begin{figure}
	\centering
	\includegraphics{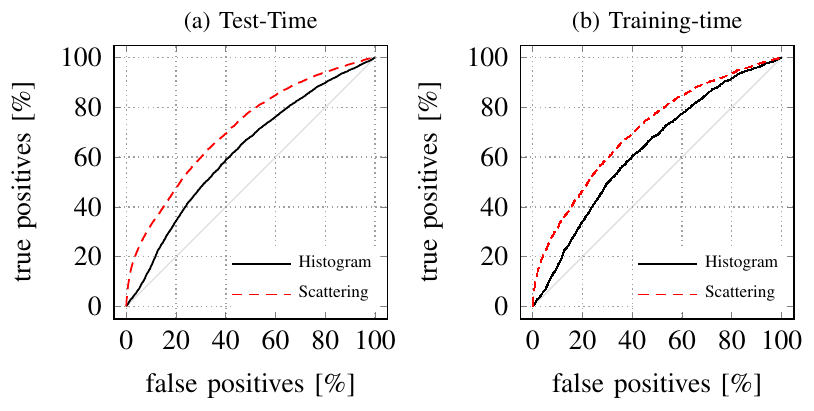}
	\vspace{-0.7em}
	\caption{Defenses against backdoor attacks: ROC curves of 
	histogram-based and color scattering-based method. Both do 
	not reliably detect image-scaling attacks that hide backdoors.}
	\label{fig:results_detection}
\end{figure}

\paragraph{Detection of image-scaling attacks.}
Another relevant question concerns the reliable detection of 
image-scaling attacks 
(see~Section~\ref{subsec:scalingattacksbackground}). 
Figure~\ref{fig:results_detection} depicts ROC curves for the 
histogram-based and color-scattering-based defense when backdoors are 
inserted at test time or training time.

For both threat scenarios, the defenses fail to detect image-scaling 
attacks. 
A closer analysis reveals that the backdoor manipulation is too small 
compared to the overall image size. Thus, down- and upscaling creates 
an image version that still corresponds to the respective input. We 
conclude that a reliable attack detection is thus not possible if small 
and bounded parts of an image are changed only.

\subsection{Clean-Label Poisoning Attack}
We proceed with the clean-label attack from
Section~\ref{subsec:poisoningwithimagescaling}, following the 
experimental setup from~\citet{ShaHuaNaj+18}. 
We test 50~randomly selected target-source class pairs $(c_t, c_z)$ 
where $c_z$ denotes the original class of $\ti$, $c_t$ the adversary's 
target class. For each pair, we choose a random target instance~\ti 
and vary the number of modified images~$X_i$. Again, a baseline is 
added where no image-scaling attack is applied on~$X_i$. For the 
embedding, we set $\alpha=0.3$.

\begin{figure}
	\centering
	\includegraphics{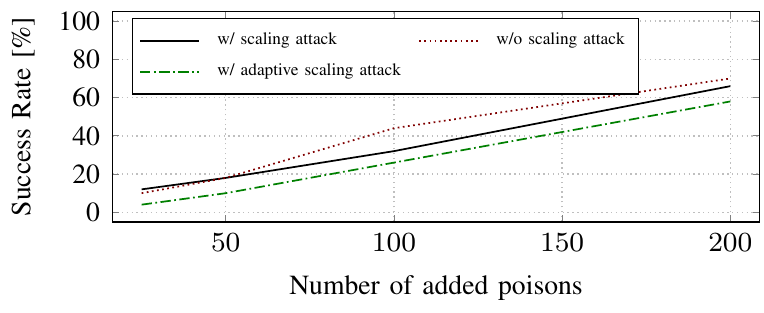}
	\vspace{-0.30cm}
	\caption{Clean-label attacks: Efficiency of attack in controlling 
	the prediction with and without image-scaling attacks, and our 
	adaptive variant.}
	\label{fig:eval_cleanlabel_scaling}
\end{figure}

\paragraph{Attack performance.}
Figure~\ref{fig:eval_cleanlabel_scaling} presents the success rate of 
the attack with respect to the number of modified images. The adversary 
can significantly control the prediction for $\ti$. The success rate 
increases with a larger number of modified images that are added to the 
training set, and corresponds to results as reported 
by~\citet{ShaHuaNaj+18}.

\begin{figure}
	\centering
	\includegraphics{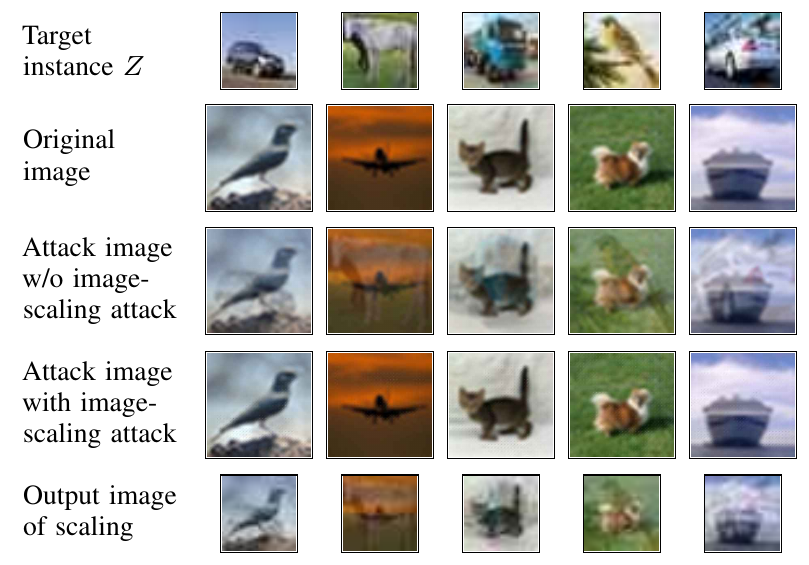}
	\caption{Clean-label attack examples. The 3rd and 4th row result in 
	the 5th row after downscaling. Image-scaling attacks can 
		effectively hide the manipulation (4th row), which would be 
		visible in the original poisoning attack~(3rd row).}
	\label{fig:eval_cleanlabel_scaling_examples}
\end{figure}

Image-scaling attacks have only a slight impact on 
the success rate of the poisoning attack. At the same time, the 
attacker can conceal the added content of $\ti$ effectively, as 
exemplified by Figure~\ref{fig:eval_cleanlabel_scaling_examples}. 
The 4th row emphasizes that the added content is not visible in the 
novel images used for training, while $\ti$ is visible for the original 
attack. %

As opposed to the backdoor attack from the previous section, the added
content \ti from the clean-label attack is not noticeable even under a
closer analysis.
As the whole image is partly changed, the manipulation becomes an 
imperceptible noise pattern. We conclude that poisoning attacks can 
benefit from image-scaling attacks the most if the manipulation is a 
weaker signal, distributed over a larger area in an image. 

\paragraph{Detection of image-scaling attacks.}
Figure~\ref{fig:results_detection_cleanlabel}(a) depicts ROC curves for
the defenses. Only the histogram-based method
can reliably detect attacks. At 1\% false positives, 94.5\% of 
manipulated images are correctly marked as attack. The 
color-scattering-based approach detects only 48.2\% at 1\% false 
positives.
In contrast to backdoor attacks, both defenses can more reliably spot 
the manipulations by the clean-label attack. As the whole image is 
slightly changed, the difference between the attack image and its down- 
and upscaled version increases---enabling the detection.

\subsection{Adaptive attack}
\begin{figure}
	\centering
	\includegraphics{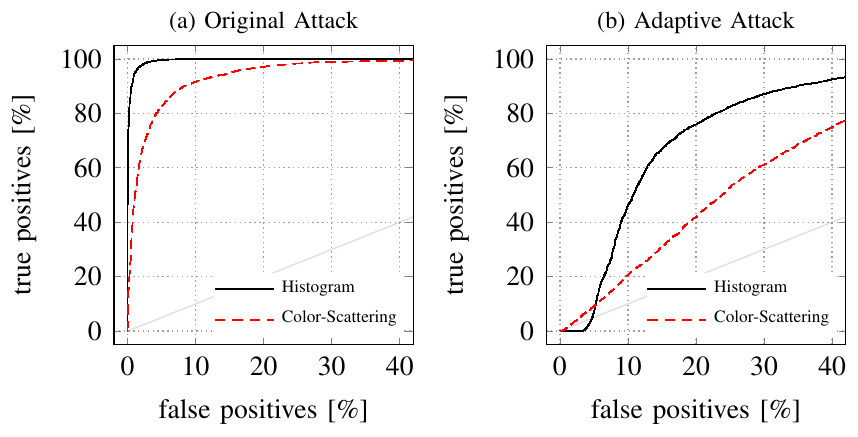}
	\vspace{-0.85em}
	\caption{Defenses against clean-label attacks: ROC curves of 
		histogram-based and color scattering-based method with the 
		original 
		and adaptive attack.}
	\label{fig:results_detection_cleanlabel}
\end{figure}

We finally demonstrate that an adversary can use an adaptive strategy 
against both defenses to lower their detection rate.
Figure~\ref{fig:results_detection_cleanlabel}(b) presents ROC curves 
for our adaptive image-scaling attack in the clean-label scenario. Our 
attack significantly lowers the detection rate. At the same time, the 
overall success rate of the attack is only slightly affected (see 
Figure~\ref{fig:eval_cleanlabel_scaling}). We contribute this to the 
histogram matching, so that parts of $\ti$ are slightly weaker 
embedded, especially for very dark or highly saturated images.
Overall, we conclude that an adversary can 
circumvent current detection methods by adjusting histograms. 

\section{Limitations}\label{sec:limitations}
Our findings demonstrate the benefit of image-scaling attacks 
for poisoning and the need to find novel detection defenses. 
Nonetheless, our analysis has limitations that we discuss in the 
following.
First, we consider defenses against image-scaling attacks only. Direct 
defenses against data poisoning~\citep[e.g.][]{WanYaoShaLi+19} are 
another possible line of defense that would need to be used after 
downscaling. The design of such defenses is an ongoing 
research problem~\citep{WanYaoShaLi+19,TanSho19} and beyond the scope 
of this work.
Furthermore, we apply a simple backdoor technique by adding a filled 
box into training images. We do not optimize regarding shapes or the 
model architecture, resulting in a relatively high amount of 
manipulated training data. Our goal is rather to draw first conclusions 
about the utility of image-scaling attacks for backdooring. As 
scaling attacks are agnostic to the model and poisoning attack, other 
backdoor techniques are also applicable whenever a manipulation needs 
to be concealed.

\section{Related Work}\label{sec:relatedwork}
The secure application of machine learning requires considering 
various attacks along a typical workflow. 
Regarding the order of the targeted step, attacks can be
categorized into the following classes:
membership inference~\citep[e.g.,][]{ShoStrSonShm17}, poisoning 
attacks~\citep[e.g.,][]{BigNelLas11, GuDolGar17, LiuMaAaf+18}, 
evasion- and perturbation
attacks~\citep[e.g.,][]{BigCorMai+13,CarWag17,QuiMaiRie19}, as
well as model stealing~\citep[e.g.,][]{TraZhaJuel+16}.

In this work, we focus on poisoning attacks that manipulate the 
training data so that the learning model returns targeted 
responses with adversarial inputs only while behaving normally for 
benign inputs. Two attack variants are backdoor and clean-label 
poisoning attacks, differing in the the amount of necessary data 
changes, visibility or robustness with transfer 
learning~\citep[e.g.][]{GuDolGar17, CheLiuLi+17, ShaHuaNaj+18, 
LiuMaAaf+18, YaoLiZhe+19}. 
We consider the following two rather simple, but representative 
approaches:
The BadNets method~\citep{GuDolGar17} inserts a small, bounded pattern 
into images as backdoor, while the clean-label attack 
from~\citet{ShaHuaNaj+18} slightly changes the whole image to add a 
poison. Both provide first insights about the 
applicability of image-scaling attacks for data poisoning.

Concurrently, \citet{QuiKleArp20} comprehensively analyze 
image-scaling attacks by identifying the root-cause and examining
defenses for \emph{prevention}. Our work here extends this line of 
research on image-scaling attacks by analyzing the poisoning 
application and \emph{detection} defenses.  While prevention stops any 
attack, detection uncovers that an attack is going on. Our findings 
here underline the need for novel detection~approaches.

\section{Conclusion}\label{sec:conclusion}
This work demonstrates that image-scaling attacks can be 
leveraged to hide data manipulations for poisoning attacks. We consider 
two representative approaches: a backdoor attack~\citep{GuDolGar17} and 
a clean-label poisoning attack~\citep{ShaHuaNaj+18}. Our evaluation 
shows that the adversary can conceal manipulations more effectively 
without impact on the overall success rate of her poisoning attack. 
We find that image-scaling attacks can create almost invisible poisoned 
instances if a slight manipulation is spread over a larger area of the 
input.

Furthermore, our work raises the need for novel detection defenses 
against image-scaling attacks. Local and bounded changes---as done for 
backdoors---are not detected at all. The detection if the whole image 
is changed can be circumvented by using our proposed adaptive 
image-scaling attack variant.

\section*{Availability}
We make our dataset and code publicly available at 
\mbox{\url{http://scaling-attacks.net}} to encourage further research 
on poisoning attacks and image-scaling attacks.

\section*{Acknowledgment}
The authors gratefully acknowledge funding by the Deutsche 
Forschungsgemeinschaft (DFG, German Research Foundation) under 
Germany's Excellence Strategy - \mbox{EXC 2092 CASA - 390781972} and 
the research grant \mbox{RI 2469/3-1}, as well as by the German 
Ministry for Education and Research as BIFOLD - Berlin Institute for 
the Foundations of Learning and Data (ref. \mbox{01IS18025A} and ref 
\mbox{01IS18037A}).

\footnotesize{
\bibliographystyle{abbrvnat}
\balance

}

\end{document}